\documentclass[useAMS,usenatbib,usegraphicx]{mn2e}
\usepackage{times} 

\newcommand{\simlt}{\lower.5ex\hbox{$\; \buildrel < \over \sim \;$}}
\newcommand{\simgt}{\lower.5ex\hbox{$\; \buildrel > \over \sim \;$}}
\newcommand{\be}{\begin{equation}}
\newcommand{\ba}{\begin{eqnarray}}
\newcommand{\ee}{\end{equation}}
\newcommand{\ea}{\end{eqnarray}}
\newcommand{\ceff}{C_{\rm eff}}
\newcommand{\bout}{B_{\rm out}}

\title[The formation history of the Galactic bulge]
	{The formation history of the Galactic bulge}
\author[I.~Ferreras, R.~F.~G.~Wyse \& J.~Silk]
{Ignacio Ferreras$^{1}$\thanks{E-mail:
ferreras@astro.ox.ac.uk}, Rosemary F.~G. Wyse$^{1,2}$ and Joseph Silk$^1$\\
$^1$Physics Dept. Denys Wilkinson Building, Keble Road, Oxford OX1 3RH\\
$^2$ The Johns Hopkins University, Dept. of Physics and Astronomy, 
Baltimore, MD 21218, U.S.A.}
\begin{document}

\date{Accepted for publication in MNRAS, 30 July, 2003}

\pagerange{\pageref{firstpage}--\pageref{lastpage}} \pubyear{2003}

\maketitle

\label{firstpage}

\begin{abstract}
The distributions of the stellar metallicities of K~giant stars 
in several fields of the Galactic bulge, taken from the literature 
and probing projected Galactocentric distances of $\sim 500$~pc 
to $\sim 3$~kpc, are compared with a simple model of star formation 
and chemical evolution. Our model assumes a Schmidt law of star formation 
and is described by only a few parameters that control the infall and 
outflow of gas and the star formation efficiency. Exploring 
a large volume of parameter space, we find that very short 
infall timescales are needed ($\simlt 0.5$~Gyr), with durations of infall 
and star formation greater than $1$~Gyr being ruled out at the 90\%
confidence level. The metallicity distributions are compatible 
with an important amount of gas and metals being ejected in outflows, 
although a detailed quantification of the ejected gas fraction is strongly 
dependent on a precise determination of the absolute stellar metallities. 
We find a systematic difference between the samples of Ibata \& Gilmore, 
at projected distances of $1-3$~kpc,  and the sample in Baade's window 
(Sadler et al.). This could be caused either by a true metallicity gradient 
in the bulge or by a systematic offset in the calibration of [Fe/H] between
these two samples. This offset does not play an important role in the 
estimate of infall and formation timescales, which are mostly dependent on the width of
the distributions. The recent bulge data from Zoccali et al. are also
analyzed, and the subsample with subsolar metallicities still rules
out infall timescales $\simgt 1$~Gyr at the 90\% confidence level.
Hence, the short timescales we derive based on  
the observed distribution of metallicities are robust and should be taken 
as stringent constraints on bulge formation models. 
\end{abstract}


\section{Introduction}
Perhaps half of the stars in the local Universe are in bulges/spheroids 
(Fukugita, Hogan \& Peebles 1998) and how and when bulges form and evolve 
are crucial clues to the origins
of the Hubble Sequence.  The determination of the star formation
history and mass assembly history of a typical bulge would provide a
stringent test of galaxy formation theories.  Plausible bulge
formation scenarios range from a single high-redshift dissipative
`star-burst' (e.g.~Elmegreen 1999), through successive
merger-induced lower-intensity star-bursts and stellar accretion
(e.g.~Kauffmann 1996), to secular models in which predominantly
stellar-dynamical effects at late epochs transform the central 
regions of thin disks to
three-dimensional bulges via bar formation and destruction
(e.g.~Raha et al. 1991; Norman, Sellwood \& Hasan 1996).  The bulge of the
Milky Way is close enough that individual stars may be studied from
the ground to within one scale-length of the centre; the combination
of deep colour-magnitude diagrams and spectroscopically determined
metallicity distributions, including elemental abundances, in
principle provides both the age distribution and the chemical
evolution.  Here we develop a model of chemical evolution of the
Galactic bulge and constrain the parameter values of our model  by
comparison with the available observational data.

There are several ways in which the bulge formation can be
traced: a spectrophotometric analysis using multi-band broadband photometry
(Peletier et al. 1999; Ellis, Abraham \& Dickinson 2001) 
or spectral indices (Proctor, Sansom \& Reid 2000) are the best approaches
for unresolved stellar populations. However, studies based on integrated 
light lack the accuracy of analyses of the properties of individual stars, 
possible for our bulge.
In this category one should mention the study of deep stellar colour-magnitude
diagrams (Feltzing \& Gilmore 2000; Zoccali et al.~2003; van Loon et al.~2003) 
and  the distribution of stellar metallicities from spectroscopic surveys 
(Ibata \& Gilmore 1995b; Sadler, Rich \& Terndrup 1996). We know from the 
analyses of deep colour-magnitude diagrams (Ortolani et al. 1995; 
Feltzing \& Gilmore 2000; Kuijken \& Rich 2002; van Loon et al.~2003)
that the vast bulk of the Galactic bulge is 'old', but the use of 
stellar isochrones cannot tell us how old to better than $\sim 3-4$~Gyr
(i.e. ages in the range $10-14$~Gyr) or tell us the dispersion of the
distribution of ages to better than that. In principle, stellar elemental
abundances constrain durations compared to lifetimes of the sources
of the elements, but these are difficult to obtain for large numbers
of stars. We demonstrate here that the shape of the overall metallicity
distribution provides a complementary constraint on the duration of
star formation, given an old age.

The available metallicity distributions for the Galactic bulge 
are reasonably well approximated by the predictions of the simple
closed-box model (e.g. Rich 1990; Zoccalli et al. 2003). In many 
applications of
this model it is a virtue that the predicted distribution is independent
of the star formation rate, but this of course precludes conclusions
about the star formation history being made by comparison of model
predictions with observations. We have thus developed a model that is 
analytically simple, but retains explicitly the timescales of gas 
flows and star formation.
Our approach makes use of the metallicity 
distributions of K~giant stars in the Galactic bulge. 
We explore a single-zone model in which no {\sl a priori}
assumptions are made regarding the parameters that describe the infall
and outflow of gas. This phenomenological ``backwards'' approach has
proven to give valuable insight on the star formation history
of early-type (Ferreras \& Silk 2000a, 2000b) as well as late-type 
galaxies (Ferreras \& Silk 2001). Our approach complements 
other studies of bulge formation, e.g. 
Moll\'a, Ferrini \& Gozzi (2000) and Matteucci, Romano \& Molaro (1999),
which assume a small set of models for a few choices of the
parameters. Instead, we explore a large volume of parameter space,
comprising thousands of model realizations which are later compared
with the observations. We are able to quantify how `closed' was the 
proto-bulge, and the timescales of inflow and star formation. 
In \S2 we describe our model; \S3 presents the data
which are used to constrain the volume of parameter space. The comparison
between model and data is discussed over the following two sections.
Finally, \S6 gives the conclusions to this work.

\section{The model}
The basic mechanisms describing star formation and the subsequent 
galactic chemical enrichment  can be reduced to the infall of primordial 
gas, metal-rich outflows, and a star formation prescription. We follow 
a one-zone model describing the star formation history in the Galactic bulge, 
which is a variation of the one used by Ferreras \& Silk (2000a; 2000b) 
to explore the formation history of early-type cluster galaxies. We define 
our model by a set of a few parameters that govern the evolution
of the stellar and gas content. A two-component system is considered, 
consisting of cold gas and stars. We adopt standard stellar lifetimes 
governing the ejection of metals but assume instantaneous mixing of the 
gas ejected by stars as well as instantaneous cooling of the hot gas 
component. We track the iron and magnesium content of the gas and the 
fraction that gets locked into stars. The basic physics related to chemical 
enrichment is reduced to:
\begin{itemize}
\item[$\bullet$] {\bf Infall:} The infall of primordial gas is described by
	a set of three parameters. We assume the infall rate -- $f(t)$ -- 
	to be an ``asymmetric'' Gaussian function with two different 
	timescales ($\tau_1$ and $\tau_2$) on either side of the peak. 
	The epoch of maximum infall is characterized by a ``formation 
	redshift'' $z_F$. Defining $\Delta t\equiv t-t(z_F)$, we can
	write the infall rate as:
	\be
	f(t) \propto \left\{ \begin{array}{cc}
	\exp [-\Delta t^2/2\tau_1^2] & \Delta t<0\\
	\exp [-\Delta t^2/2\tau_2^2] & \Delta t\geq 0\\
	\end{array}
	\right.
	\ee
\item[$\bullet$] {\bf Outflows:} Outflows of gas and metals triggered 
	by supernova explosions constitute another important factor 
	contributing to the final metallicity of the bulge. We define 
	a parameter $\bout$ which represents the fraction of gas and
	metals ejected from the galaxy. Even though one could
	estimate a (model-dependent) value of $\bout$ given the star 
	formation rate and 
	the potential well given by the total mass of the galaxy
	(e.g. Larson 1974, Arimoto \& Yoshii 1987), we leave $\bout$
	as a free parameter.
\item[$\bullet$] {\bf Star Formation Efficiency:} Star formation is
  assumed to follow a variant of the Schmidt law (Schmidt 1959): 
  \be \psi (t)=\ceff
  \rho_g^{1.5}(t), \label{eq:schmidt} 
  \ee 
  where $\rho_g$ is the gas
  volume density, and the parameter $\ceff$ gives the star formation
  efficiency, which -- for a linear Schmidt law -- is an inverse
  timescale for the processing of gas into stars. The $1.5$ exponent was
  chosen based on the best fit to observations from a local sample of
  normal spiral galaxies (Kennicutt 1998). This exponent is also what
  one obtains for a star formation law that varies linearly with gas 
  gas density and inversely with  the local dynamical
  time, since for self-gravitating gas disks, this timescale varies 
  as the inverse square root of the gas density. 
  Changing the slope of this star formation dependence on gas density  will
  alter the inferred timescales of star formation;  other
  slopes for this correlation are explored in \S5.3.
\end{itemize}

The input parameters are thereby five: $(\tau_1,\tau_2,\bout,$
$\ceff ,z_F)$. The equations can be separated into one set 
that follows the mass evolution and another set that traces chemical 
enrichment. The evolution of the mass in gas and stars is given by:
\begin{equation}
{d\rho_g\over dt} = (1-\bout)E(t)-\psi (t)+f(t)
\end{equation}
\begin{equation}
{d\rho_s\over dt} = \psi (t) - E(t)
\end{equation}
\begin{equation}
E(t) = \int_{M_t}^\infty dM\phi (M)(M-w_M)\psi (t-\tau_M),
\end{equation}
where $\phi(M)$ is the initial mass function (IMF). 
The integral $E(t)$ is the gas density ejected at time $t$ from 
stars which have reached the end of their lifetimes. $\tau_M$ is the 
lifetime of a star with mass $M$. 
We describe stellar lifetimes as a broken power law fit 
to the data from Tinsley (1980) and Schaller et al. (1992):
\begin{equation}
\left( {\tau_M\over { Gyr}} \right) = \left\{
\begin{array}{lr}
 9.694  \left( {M\over M_\odot}\right)^{-2.762} & M < 10M_\odot\\
 0.095  \left( {M\over M_\odot}\right)^{-0.764} & M > 10M_\odot\\
\end{array}
\right.
\end{equation}
$M_t$ is the turnoff mass, i.e.~the mass of a main sequence star which reaches
the end of its lifetime at a time $t$. Finally, $w_M$ is the stellar remnant 
mass for a star with main sequence mass $M$:
\begin{equation}
\left( w_M\over M_\odot \right) = \left\{
\begin{array}{ll} 
 0.14(M/M_\odot )+0.36   & M/M_\odot \leq 8\\
 1.5                    & 8<(M/M_\odot)\leq 25\\
 0.61(M/M_\odot )-13.75 & M/M_\odot >25\\
\end{array}
\right.
\end{equation}

The mass for white dwarf remnants was taken from Iben \& Tutukov (1984). 
The $1.5M_\odot$ remnant mass given for 
the intermediate range is the average mass of a neutron star
(e.g. Shapiro \& Teukolsky 1983), whereas supernovae 
from heavier stars might give birth to black holes, locking more 
mass into remnants (Woosley \& Weaver 1995). The equations for the 
evolution of the metallicity of the gas and the stars 
are  -- see Ferreras \& Silk (2000a; 2000b) for details:
\begin{equation}
\begin{array}{ll}
d(Z_g\rho_g)/dt& =-Z_g(t)\psi (t)+(1-\bout)E_Z(t)\\
\end{array}
\end{equation}
\begin{equation}
\begin{array}{ll}
d(Z_s\rho_s)/dt& = Z_g(t)\psi (t)-\int_{M_t}^\infty 
dM\phi (M) \\
 & \\
 &\times (M-w_M-Mp_M)(Z_g\psi )(t-\tau_M)\\
\end{array}
\end{equation}
\begin{equation}
\begin{array}{ll}
E_Z(t)& =\int_{M_t}^\infty dM\phi (M)\big[ (M-w_M\\
 & \\
 & -Mp_M)(Z_g\psi )(t-\tau_M)+Mp_M\psi (t-\tau_M)\big].\\
\end{array}
\end{equation}

The yields -- $p_M$ -- are defined as the fraction of 
a star of mass $M$ transformed into metals and ejected into the ISM. 
The model tracks the evolution of Mg and Fe with the yields of
Thielemann, Nomoto \& Hashimoto (1996) for core-collapse supernovae 
(solar metallicity progenitors) and from model W7 in Iwamoto 
et al. (1999) for type~Ia supernovae.

\begin{figure}
\includegraphics[width=3.4in]{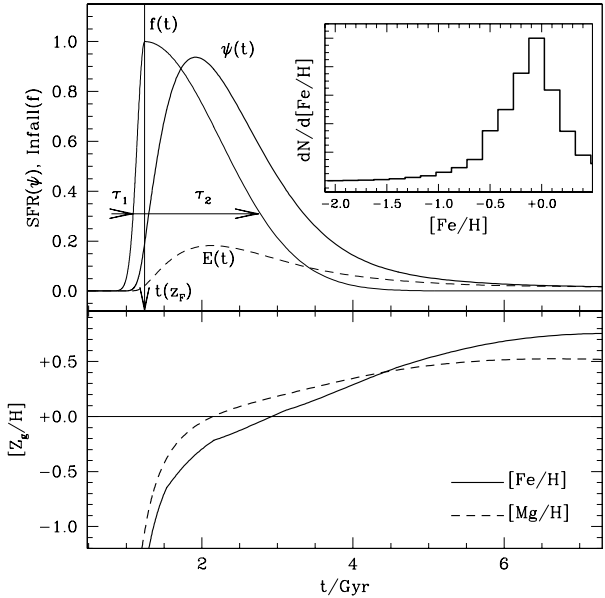}
\caption{Star formation history given by the model for $\tau_1=0.1$~Gyr; 
$\tau_2=1$~Gyr; $\ceff = 5$; $z_F=5$; $\bout =0.2$. The top panel
shows the star formation rate ($\psi$, thick line), gas infall rate
($f$, thin) and gas ejected from stars ($E$, dashed).  The bottom
panel gives the time evolution of [Mg/H] (dashed) and [Fe/H] (solid)
in the gas, and thus in stars formed at that time. The inset in
the top panel is the final histogram of true iron metallicities in
long-lived stars, to be compared with the data for the bulge fields
described in the text.}
\label{fig:sfh}
\end{figure}
  
In order to estimate the rates of type~Ia supernovae (SNeIa), we 
follow the prescription of Greggio \& Renzini (1983), recently 
reviewed by  Matteucci \& Recchi (2001). We assume the progenitor 
of each SNeIa is a single degenerate close binary system in which
a CO white dwarf accretes gas from the non-degenerate companion
triggering a carbon deflagration. The rate of SNeIa can be written as 
a convolution of the IMF over the mass range that can generate such
a binary system. We assume a lower mass limit of $M_{Bm}=3M_\odot$ 
in order to have a binary with a CO white dwarf which will reach the 
Chandrasekhar limit after accretion from the secondary star 
(Matteucci \& Greggio 1986). Notice that this limit is somewhat
uncertain and will be lower if He white dwarfs can give rise
to SNeIa (Greggio \& Renzini 1983). The upper mass limit is $M_{BM}=16M_\odot$
so that neither binary undergoes core collapse (assumed to happen in
stars more massive than $8M_\odot$). 
The rate is thus:
\be
R_{Ia}(t)={\cal A}\int_{3M_\odot}^{16M_\odot}dM\phi (M)
\int_{\mu_M}^{\mu_{\rm max}}d\mu^\prime f(\mu^\prime )\psi (t-\tau_{M_2}),
\label{eq:sn1a}
\ee
where $\tau_{M_2}$ is the lifetime of the nondegenerate companion, 
with mass $M_2$. $f(\mu )$ is the fraction of binaries
with a mass fraction $\mu\equiv M_2/M_B$, where $M_2$ and $M_B$
are the masses of the secondary star and the binary system,
respectively. The range of integration goes from 
$\mu_M={\rm max}(M_2/M,1-8M_\odot/M)$, to $\mu_{max}=0.5$. 
The analysis of Tutukov \& Yungelson (1980) on a sample of about 1000
spectroscopic binary stars suggests that mass ratios close
to $\mu=1/2$ are preferred, so that the normalized distribution 
function of binaries can be written: 
\be
f(\mu ) = 2^{1+\gamma}(1+\gamma )\mu^\gamma, 
\ee
as suggested by Greggio \& Renzini (1983), and we adopt the value 
of $\gamma =2$. The normalization constant ${\cal A}$ 
(equation~\ref{eq:sn1a}) is 
constrained by the ratio between type~Ia and type~II supernovae 
in the solar neighbourhood 
that best fits the observed solar abundances. We use the
result of Nomoto, Iwamoto \& Kishimoto (1997), namely
$R_{Ia}/R_{II}=0.12$ to find ${\cal A}=0.05$, although there
is still a rather large uncertainty in the ratio of supernova 
rates, so that $R_{Ia}/R_{II}$ can be as high as $0.3$
(e.g. Iwamoto et al. 1999), which would imply ${\cal A}\sim 0.12$.

We want to emphasize that the model presented here is not a variation
of the Simple Model (e.g. Pagel 1997), as we assume that all of the
gas in the system comes from infall as described above and the
Simple Model usually incorporates the Instantaneous Recycling
Approximation, which we do not. Our model reduces to a Simple Model
either in the limit $\tau_1,\tau_2\rightarrow 0$ or 
when we set the infall rate $f(t)=0$ and 
assume a non-zero initial gas content. Furthermore, the
addition of SNeIa implies the presence of more high-metallicity stars
(with lower [$\alpha$/Fe] abundance ratios) with respect to the Simple
Model, if the duration of the star formation history is comparable to
the onset time for type~Ia supernovae. We do not assume any
proportionality between infall or outflows and the star formation
rate, in contrast with, e.g. Hartwick (1976) or Mould (1984). Our
model is complementary to other work on chemical enrichment in the
bulge (e.g. Moll\'a, Ferrini \& Gozzi 2000; Matteucci, Romano
\& Molaro 1999) in the sense that we let all the parameters
controlling the SFH to vary over a wide range, only constraining this
parameter space with the observations.  Figure~\ref{fig:sfh} shows the
star formation history obtained for a choice of parameters:
($\tau_1=0.1$~Gyr,$\tau_2=1$~Gyr, $\bout =0.2$ ,$\ceff
=5$,$z_F=2$). The top panel shows the infall rate ($f$) the star
formation rate ($\psi$), and the gas ejected from stars ($E$).  The
lag between $f(t)$ and $\psi (t)$ is caused by a combination of
factors: a finite star formation efficiency, the contribution of gas
from stars ($E$), and the power law used to describe the star
formation rate as a function of gas density
(equation~\ref{eq:schmidt}). The infall parameters $\tau_1$,
$\tau_2$, and $z_F$ are shown in the figure. The bottom panel gives
the evolution of [Mg/H] (dashed) and [Fe/H] (solid) as a function of
age.  One can see Mg dominates at the early stages, whereas the higher
Fe yields from SNeIa make the ISM more iron rich at late times. The
inset in the top panel is the metallicity histogram of the
simulation. 
Infall of metal-poor gas provides narrower metallicity 
distributions by producing relatively more stars at later times compared 
with no-infall models (such as the Simple Model). In extreme cases
the infall rate can be balanced with the star formation rate 
so that the metallicity presents very narrow distributions
(Lynden-Bell 1975).
The low metallicity tail of the histogram is mainly
``driven'' by the early infall timescale $\tau_1$, which controls the
buildup of metals from the primordial infalling gas.
The second infall timescale $\tau_2$ has a stronger
effect on the high metallicity part of the histogram since it only
controls the infall of gas after the epoch of maximum infall, i.e.
once the average metallicity of the ISM is rather high.
The outflow parameter $\bout$ also plays a
very important role in ``modulating'' the net yield, so that an
increase in $\bout$ shifts the histogram towards lower metallicities
(cf.~Hartwick 1976).  Although only weakly, outflows can also
contribute to the shape of the histogram since the duration of the
star formation stage is shortened when high outflow fractions are
considered, Hence, we decide to freely explore these three parameters
($\tau_1$,$\tau_2$,$\bout$) and to fix the star formation efficiency
and formation epoch to be compared for a few realizations.  We focused
on three models defined by ($\ceff =10$,$z_F=5$), ($\ceff
=5$,$z_F=5$), and ($\ceff =10$,$z_F=2$). The choice is justified by
the expected low fraction of young stars observed in the bulge, which
suggests high star formation efficiencies and early formation
epochs. We choose a $\Lambda$CDM cosmology with $\Omega_m=0.3$;
$\Omega_\Lambda =0.7$; $H_0=65$~km/s/Mpc, which implies infall maxima
ocurring at around $1.2$ and $3.5$~Gyr for formation redshifts $z_F=5$
and $2$, respectively. The age of the Universe for this cosmology is
$14.5$~Gyr.

\begin{figure}
\includegraphics[width=3.4in]{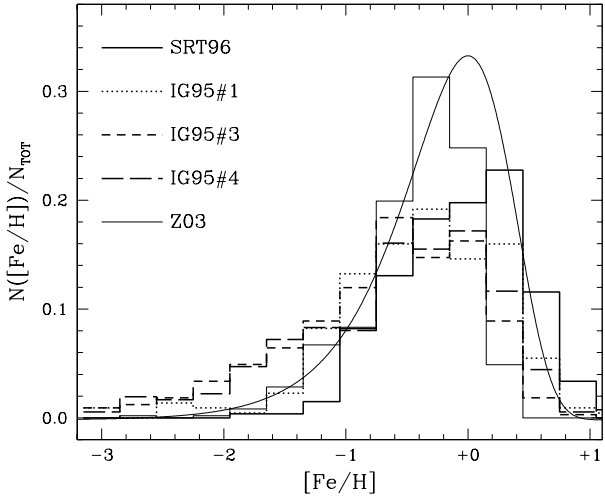}
\caption{Iron abundance histograms of the  stars in the bulge fields used 
in this paper. Notice the offset in the  distributions between the IG95 
fields and SRT96 and Z03; a  systematic offset of $\sim 0.3$~dex would 
``align'' the histograms of these fields. The iron abundances are all 
obtained using different techniques, and the offsets could reflect an 
underlying difference in calibrations, as discussed in the text.
The curve gives the expected distribution for the Simple Model, with a
yield $y=Z_\odot$.}
\label{fig:data}
\end{figure}

\section{The data}
We use the derived chemical abundance distributions from observations
of $K$~giants in various fields towards the Galactic bulge from two
samples: Ibata \& Gilmore (1995a,b; hereafter IG95) as well as Sadler,
Rich \& Terndrup (1996; hereafter SRT96).  K-giants are the
preferred tracer since they are relatively unbiased with respect to
age or metallicity. We also consider the recent observations of the
bulge from Zoccali et al. (2003; hereafter Z03). 
IG95 target several fields towards the
Galactic bulge along the minor and major axes, at projected
Galactocentric distances of $\sim 1-3$~kpc, to mimic the long-slit
spectroscopy of external bulges, whereas SRT96 concentrate on Baade's
window, i.e. $(l,b)=(1\degr ,-3.9\degr )$, with the Z03 field
somewhat further in projected Galactocentric distance (see Table~1).

We have restricted our consideration to those fields in which a
significantly large number of stars have derived metallicities, 
indeed iron abundances.  We should emphasize that the [Fe/H]
metallicities which are used to compare against our model have been
obtained using different techniques.  IG95 compute the `iron'
metallicity from a metallicity-sensitive Mg-spectral index (around
5200\AA), calibrated together with $B-V$ colour (sensitive to stellar
effective temperature) onto a [Fe/H] scale using a sample of K giant
stars in the solar neighbourhood (Faber et al. 1985).  Thus their
technique implicitly assumes that the programme stars in the bulge and
the local standard stars have the same value of [Mg/Fe] at a given
[Fe/H].  As they noted, and we return to this point below, this may
well not be the case, perhaps necessitating a new calibration of iron
for their sample.

On the other hand, SRT96 estimate [Fe/H] from Fe spectral lines plus
$V-I$ colour (again the colours are needed in order to include
sensitivity to stellar effective temperatures). The calibration stars
are the same sample from Faber et al. (1985). Hence, a systematic and
possibly non-linear offset between these two samples could be
expected, given the inference from high-resolution studies of a very
limited sample, that many of the stars in Baade's window have enhanced
magnesium, with [Mg/Fe] $\sim +0.3$ (McWilliam \& Rich 1994; see also
Maraston et al.~2002).  Finally, Z03 perform a metallicity estimate
using only photometric data. After a careful removal of disk stars,
the metallicities are computed by a comparison of the locus of RGB
stars in the ($M_K$ vs $V-K$) colour-magnitude diagram with
analytical representations of RGB templates from Galactic globular
clusters over a range of metallicities and thus an old age is
assumed implicitly.  As the authors caution, this analysis may
suffer from a strong systematic effect and the expected error bars
should be larger than metallicities obtained with spectroscopy.  
The authors further obtain an iron abundance distribution from their
metallicity distribution by subtracting an $\alpha$-element
enhancement of 0.2~dex for stars with [Fe/H]$> -1$ and 0.3~dex for
more iron-poor stars.


\begin{table}
\caption{Bulge Fields}
\label{tab:sample}
\begin{center}
\begin{tabular}{cc|ccc}
FIELD & ($l,b$) & N & N$_{\rm sub}$ & M([Fe/H])$_{\rm sub}$\\
\hline\hline
SRT96    &  ($+1\degr$,$-3.9\degr$) & $268$ & $140$ & $-0.37$\\
IG95/\#1 &  ($-25\degr$,$-12\degr$) & $219$ & $153$ & $-0.57$\\
IG95/\#3 &  ($-5\degr$,$-12\degr$)  & $326$ & $269$ & $-0.71$\\
IG95/\#4 &  ($+5\degr$,$-12\degr$) & $361$ & $271$ & $-0.66$\\
 Z03     &  ($+0.3\degr$,$-6.2\degr$) & $503$ & $360$ & $-0.38$\\
\end{tabular}
\end{center}
\end{table}

\subsection{The systematics of abundance measurements}

  Table~\ref{tab:sample} shows the fields explored in this paper along
with the number of bulge stars observed ($N$) and the number of those
stars with subsolar metallicities ($N_{\rm sub}$).  The last column
gives the median value of [Fe/H] for the subsolar sample.
Figure~\ref{fig:data} shows the metallicity histogram of all fields
considered in this paper. A significant offset between the IG95 fields
and SRT96 can be readily seen. The prediction for a Simple Model of
chemical enrichment with a net yield $y=Z_\odot$ is shown as a solid
line; for solar elemental ratios this corresponds to [Fe/H]$_\odot$.
The total yield is defined as:
\be
y = \frac{1}{1-R}\int_{M_1}^\infty dMMp_M\phi (M),
\ee
where $M_1$ is the present turnoff mass ($\sim 1M_\odot$) and $R$ is the
returned fraction, i.e. the amount of gas returned from stars once they
reach their endpoints, namely
\be 
R = \int_{M_1}^\infty dM(M-w_M)\phi (M).
\ee
All histograms would peak at similar metallicities if the 
IG95 fields were shifted towards higher metallicities by $\sim +0.3$~dex.  
Notice that both SRT96 and Z03 have a similar value of
M([Fe/H])$_{\rm sub}$.  Even though these two fields are not too far
from each other compared to the IG95 fields, they are still separated
a projected distance of 300~pc, i.e. around a scalelength of the
bulge. The full histogram of SRT96 and Z03 is significantly different.

The offset between IG95 and SRT96 can be a systematic effect since
the estimate of [Fe/H] from Mg spectral indices and $B-V$ colour uses
stars with solar abundance ratios as calibrators. This is a perfectly
valid method for local stars.  However, we would expect stars with
enhanced [Mg/Fe] -- such as established for a small sample of stars in
the Galactic bulge (McWilliam \& Rich 1994) -- to give systematically
lower [Fe/H] if the same calibration stars are used. The sample of
SRT96 use Fe spectral lines which should give a better approximation
to [Fe/H] even if calibrators with solar abundance ratios are used.
We can roughly estimate how much of an offset would be expected
between these two data sets using the correction to the metallicity
suggested by Salaris, Chieffi \& Straniero (1993). The isochrones
for solar abundances corresponding to the corrected metallicity
are equivalent to those for enhanced [Mg/Fe] isochrones taking 
the original value of the metallicity. Assuming an enhancement 
of [Mg/Fe]=+0.3~dex (Rich \& McWilliam 2000), the correction
to the metallicity obtained when the calibration stars have solar
[Mg/Fe] would result in an offset of 0.21~dex,
which is more or less the observed discrepancy between the median
[Fe/H] between the IG95 fields and SRT96.

On the other hand, this `discrepancy' could be due to real 
astrophysics, as expected if the amount of metals
driven by outflows had a strong dependence on the radial position
within the bulge. Table~\ref{tab:sample} shows that all three IG95 fields
explored in this paper are at a Galactic latitude of $-12\degr$,
whereas Baade's window is found at $b=-3.9\degr$, i.e. the IG95 fields
are at a projected distance $\simgt 1.1$~kpc away from the centre,
assuming a distance of $7.9$~kpc to the Galactic centre (McNamara et
al. 2000). The upshot is that we believe the conclusions from our
analysis are very robust regarding infall timescales -- which pertain
to {\sl relative} metallicities -- whereas the estimated fraction of
gas ejected in outflows -- which depend very sensitively on {\sl
absolute} metallicities -- should be taken with care, given the
possible systematic offsets between the different studies used. 

However, throughout this paper we shall treat all [Fe/H] estimates
on an equal basis and accept them at face value.

\begin{figure}
\includegraphics[height=3.3in,width=3.3in]{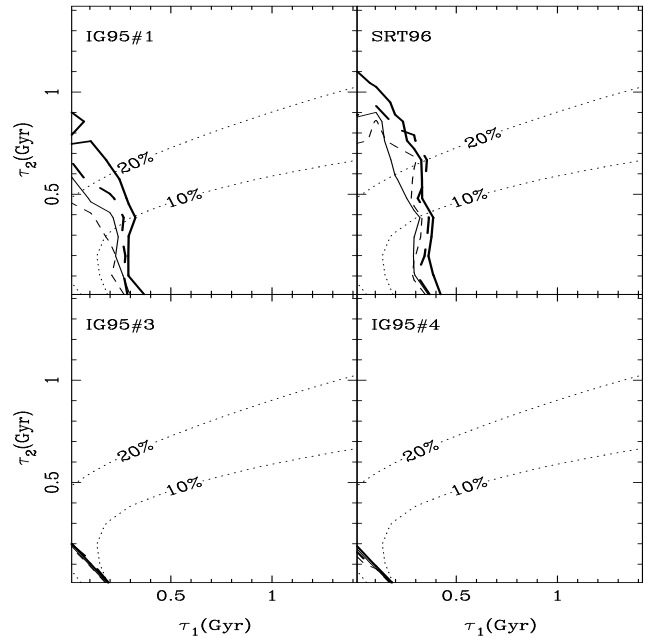}
\caption{Probability maps at the $90\%$ and $95\%$ (thick)  
confidence levels in the ($\tau_1$,$\tau_2$) parameter space for the
$\ceff =10$,$z_F=5$ (solid) and $\ceff =10$, $z_F=2$ (dashed) models. 
The dotted lines give the fraction of stars younger than 10~Gyr. The panels 
show the analysis for all the fields listed in table~\ref{tab:sample}.
The analysis is performed with the sample comprising subsolar 
metallicities.}
\label{fig:t1t2}
\end{figure}

\section{Comparing model and data}
We compare the samples discussed above with our model prediction
by performing a Kolmogorov-Smirnov (KS) test. The computational method
is rather intensive. 
Furthermore, there is a degeneracy between infall 
timescale and star formation efficiency so that the width of the
metallicity distribution  depends on the ratio of these 
two parameters (see e.g. Lynden-Bell 1975).  
We can break this degeneracy by using the independent evidence of 
the age distribution in the Galactic bulge, specifically the low 
fraction of young and intermediate-age stars inferred from 
deep colour-magnitude diagrams 
(Feltzing \& Gilmore 2000; Kuijken \& Rich 2002; 
van Loon et al.~2003).  We can thus 
assume a fixed (and high)
value of $\ceff$, leaving the infall timescale as a free parameter.
Hence, we decided to fix the star formation efficiency ($\ceff$) and formation 
epoch ($z_F$) for three models, namely ($\ceff =10$,$z_F=5$), 
($\ceff =5$,$z_F=5$), and ($\ceff =10$,$z_F=2$). 
The remaining three parameters are explored over a wide range: 
$0.01 < \tau_{1,2}/{\rm Gyr} < 1.5$; $0.0 < \bout < 0.8.$. 
Models with $\ceff =10$; $\tau_1=\tau_2=0.5$~Gyr
form 50\% of the stars in $\sim 0.75$~Gyr, compared to a 
longer duration of $\sim 1.9$~Gyr when the efficiency is
lowered to $\ceff=1$.

We computed a grid of $16\times 16\times 16$ star formation 
histories and compared the simulated metallicity histograms with
the data described above. The calibration of stars with inferred 
supersolar iron  abundances is rather complicated and implies large 
uncertainties. Hence, two subsamples  of the data were used for each field: 
one in which all observed stars were included and a second set
in which only stars with subsolar iron abundances were included 
in the histograms. We obtained a final 3D array with each 
element representing the KS probabilities for a given choice 
of parameters $(\tau_1,\tau_2,\bout )$. Table~\ref{tab:sub}
gives the 90\% confidence levels for these parameters, using 
only stars with subsolar iron abundances. The results obtained 
in a comparison with the full sample is shown in 
table~\ref{tab:all}. A KS test is an optimal statistic to be
used with unbinned data. However, given that different statistical
tests are sensitive to different properties of the distribution, we
also performed a $\chi^2$ test on binned samples. This test is rather
dependent on binning. However, the number of stars in each sample is
large enough (table~\ref{tab:sample}) to make the comparison worthwhile.
We computed the $\chi^2$ by taking 10 bins in the range $-3<[$Fe/H$]<+1$
in the metallicity distributions of both the observed data and the models.
The parameters we obtained using a $\chi^2$ test were very similar
to those shown in tables~\ref{tab:sub} and \ref{tab:all} and always
within the 95\% confidence levels. Hence, we consider the best 
fit parameters obtained to be rather insensitive to the statistical 
estimator used.

\begin{figure}
\includegraphics[height=3.3in,width=3.3in]{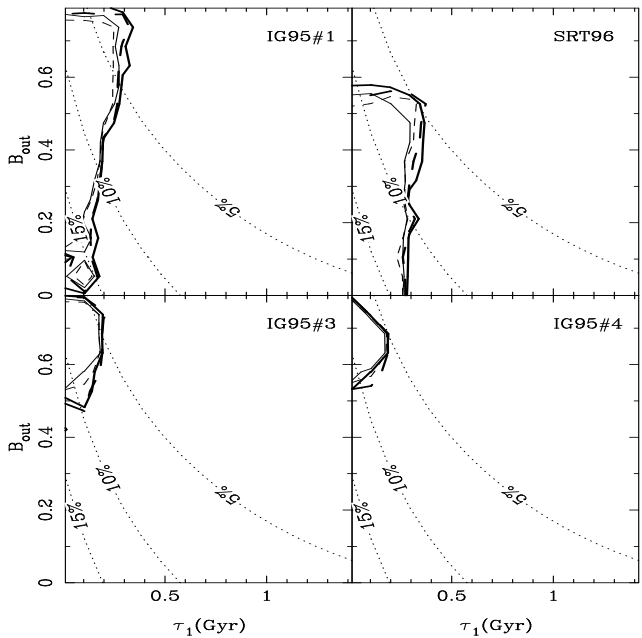}
\caption{Probability maps at the $90\%$ and $95\%$ confidence levels in the
($\tau_1$,$\bout$) parameter space. The same notation as in 
Figure~\ref{fig:t1t2} is used.}
\label{fig:t1b}
\end{figure}

Figures~\ref{fig:t1t2},\ref{fig:t1b}, and \ref{fig:t2b} give probability maps 
of the KS test at the 90\% (thin) and 95\% (thick) confidence levels.
Two models are given: $\ceff =10$, $z_F=5$ (solid) and
$\ceff =10$, $z_F=2$ (dashed). Each figure shows a 2D projection
of the three dimensional volume spanned by ($\tau_1$,$\tau_2$,$\bout$).
Figure~\ref{fig:t1t2} shows that the model is incompatible
with infall timescales longer than $\sim 1$~Gyr. Given the high star 
formation efficiencies used in the model, this translates into a
very short duration of the star formation stage.  
The results are
rather insensitive to the formation epoch chosen. Tables~\ref{tab:sub}
and \ref{tab:all} show that the best fit parameters are roughly the
same for $z_F=5$ or $z_F=2$ as long as the same star formation efficiency
is chosen. This reflects that fact that the stellar metallicity distribution
is mostly sensitive to the duration of star formation and does not
depend much on absolute ages. A lower star formation efficiency
results in longer infall timescales, as seen in the tables.
These timescales are still shorter than 1~Gyr at the 90\% confidence
level  for the more reliable subsolar metallicities, 
except for field SRT96.

The available colour-magnitude diagrams for lines-of-sight in the
bulge ranging from projected distances from the Galactic 
Center of $\sim 100$~pc to many kpc, imply that the vast
bulk of the bulge stars are old, with ages greater than $\sim 10$~Gyr
(Ortolani et al. 1995; Feltzing \& Gilmore 2000; Kuijken \& Rich 2002; 
van Loon et al.~2003).  
We incorporate these results into the comparison
between models and data in terms of a simple fraction of stars
predicted to be younger than 10~Gyr, shown as dotted lines in the 
figures. These mass fractions are computed for a model with
$\ceff = 10$; $z_F=2$ and, where applicable, the other parameters
are fixed to $\bout =0$ (figure~\ref{fig:t1t2}); $\tau_2=0.5$~Gyr 
(figure~\ref{fig:t1b}) or $\tau_1=0.5$~Gyr (figure~\ref{fig:t2b}).
These parameters were chosen to represent a conservative scenario
in the estimates of young stellar mass fractions. Hence, the figures
show that the assumption of a stellar mass fraction no greater
than 20\% in stars younger than 10~Gyr imposes a further constraint 
on the infall timescales, so that $\tau_2\simgt 0.5$~Gyr would be 
ruled out. The estimates for the infall timescales shown in the 
tables do not take this constraint into account. However, it is 
worth pointing this out as an additional factor favouring very short
infall timescales.


\begin{table}
\caption{Best fit parameters ([Fe/H]$<0$; 90\% confidence levels)}
\label{tab:sub}
\begin{center}
\begin{tabular}{c|cccc}
FIELD & Model & $\tau_1$/Gyr & $\tau_2$/Gyr & $\bout$\\
\hline\hline
IG95/\#1 & C10z5 & $<0.19$ & $<0.51$ & $0.48^{+0.24}_{-0.34}$\\
         &  C5z5 & $<0.40$ & $<0.79$ & $0.43^{+0.29}_{-0.21}$\\
         & C10z2 & $<0.17$ & $<0.47$ & $0.48^{+0.23}_{-0.33}$\\

SRT96    & C10z5 & $<0.26$ & $<0.75$ & $0.11^{+0.39}_{-0.11}$\\
         &  C5z5 & $<0.79$ & $<1.03$ & $0.05^{+0.45}_{-0.05}$\\
         & C10z2 & $<0.25$ & $<0.74$ & $0.32^{+0.17}_{-0.32}$\\

IG95/\#3 & C10z5 & $<0.09$ & $<0.10$ & $0.69^{+0.04}_{-0.17}$\\
         &  C5z5 & $<0.15$ & $<0.18$ & $0.64^{+0.10}_{-0.10}$\\
         & C10z2 & $<0.10$ & $<0.10$ & $0.69^{+0.04}_{-0.18}$\\

IG95/\#4 & C10z5 & $<0.10$ & $<0.10$ & $0.69^{+0.04}_{-0.15}$\\
         &  C5z5 & $<0.10$ & $<0.10$ & $0.69^{+0.04}_{-0.15}$\\
         & C10z2 & $<0.09$ & $<0.08$ & $0.69^{+0.03}_{-0.15}$\\

Z03      & C10z5 & $<0.16$ & $<0.33$ & $0.16^{+0.21}_{-0.16}$\\
         &  C5z5 & $<0.24$ & $<0.51$ & $0.05^{+0.35}_{-0.05}$\\
         & C10z2 & $<0.17$ & $<0.37$ & $0.05^{+0.41}_{-0.05}$\\
\end{tabular}
\end{center}
\end{table}

\section{Discussion}

\subsection{The formation history of the bulge}

\begin{figure}
\includegraphics[width=3.5in]{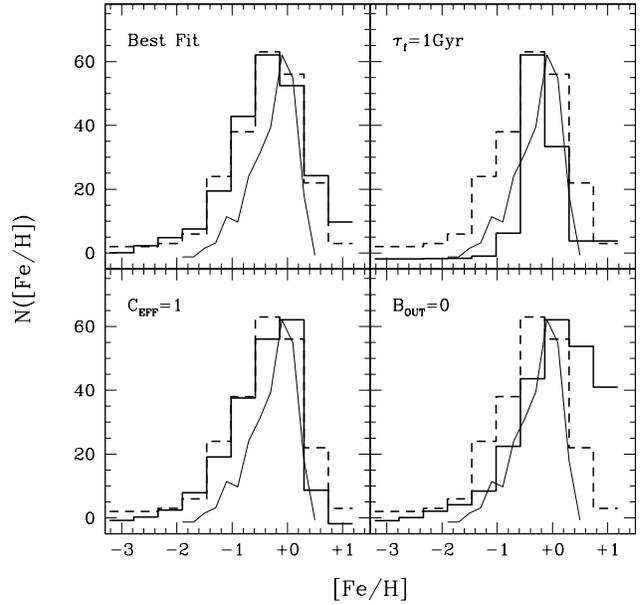}
\caption{A comparison of the effect of varying the parameters explored
in this paper. The dashed line in all panels give the metallicity
distribution of the sample in Ibata \& Gilmore field IG95\#1. The solid
lines are the model predictions. The top-left panel corresponds to 
$\tau_1=\tau_2=0.05$~Gyr; $\bout=0.5$; $\ceff =10$; $z_F=5$, which gives 
a good fit to the observed data (KS test probability $94\%$). 
The remaining three panels show the predicted histograms when 
changing one of the parameters to the value shown keeping the other
parameters fixed. In the top=right panel, $\tau_f$ represents the
overall infall timescale $\tau_1+\tau_2$. Notice infall timescales 
of $1$~Gyr are readily ruled out because of the narrow metallicity
distributions which are generated.  The thin line in all 
four panels give the metallicity distribution of field Z03, which
is narrower than IG95\#1 thereby allowing for a slightly longer
infall timescale.
}
\label{fig:hist}
\end{figure}

Figure~\ref{fig:hist} explores the effect of varying the parameters 
used in this paper, on the stellar metallicity distribution. In all
panels, the dashed line gives the distribution of IG95 field \#1. 
The solid lines are model predictions.
The {\sl top, left} panel gives the best fit from our model to the data,
corresponding to $\tau_1=\tau_2=0.05$~Gyr; $\bout =0.5$; 
$\ceff =10$ (for a fixed $z_F=5$). The remaining three panels show the
predicted histograms when varying any of the parameters explored in 
this paper, keeping all other parameters fixed to the best fit values.
We also show -- as a comparison -- the histogram of the Z03 field.
In the {\sl bottom, right} panel, the fraction of gas and metals ejected 
in outflows is changed to $\bout =0$ (keeping all other parameters 
unchanged). The 
resulting histogram corresponds -- to first order -- to an offset
towards higher metallicities, since more gas is allowed to be locked
into subsequent generations of stars. The shape of the histogram will also be
slightly modified since a low value for the outflow fraction 
allows a longer duration of star formation. 
On the other hand, changing the stellar yields would 
mimic a change in $\bout$, so that given the uncertainties in 
the Fe yields from simulations of supernova explosions 
(Woosley \& Weaver 1995; Thielemann et al. 1996; Iwamoto et al. 1999) 
we can conclude that the absolute determination of $\bout$ may still 
carry an important systematic offset. The {\sl bottom, left} panel of 
figure~\ref{fig:hist} shows the effect of a lower star formation 
efficiency. The histogram does not change much --  to be expected 
given the very short infall timescale considered -- although the lower
$\ceff$ tends to give lower metallicities. 
We will show below
that these models can also be discriminated if [Mg/Fe] abundance
ratios are used in the analysis. 

The shape of the metallicity distribution is strongly affected by 
a change in the infall timescale (cf.~the infall solution to the 
local disk `G-dwarf problem', Tinsley 1975). 
Instead of varying $\tau_1$ and $\tau_2$ separately, 
we show on the {\sl top, right} panel the effect of extending the
total infall timescale $\tau_f=\tau_1+\tau_2$ to 1~Gyr. 
A more extended infall results in a
higher fraction of stars with higher [Fe/H], thereby sharpening
the metallicity distribution. The prominent tail of the histogram
observed at low metallicities shows that long infall timescales
are not allowed by the observations. Quantitatively, 
figure~\ref{fig:hist} shows that star formation
timescales longer than 1~Gyr are unlikely. 
Notice that field Z03 features a narrower distribution of
metallicities, thereby favouring longer infall timescales. However,
as shown in table~\ref{tab:sub} the analysis of the (more reliable)
sample comprising stars with subsolar metallicities still discard
infall timescales $\tau_f\simgt 1$~Gyr for field Z03 at 
the 90\% confidence level. It is also worth remembering  that
these theoretical distributions assume a very high star formation
efficiency -- as expected from the lack of young stars in the
Galactic bulge. A lower value of $\ceff$ will imply a wider
distribution of metallicities as $\tau_f$ is increased. It is only
the case of a very high $\ceff$ along with the assumption
of instantaneous mixing that  gives narrow distributions
when extended infall is assumed.

\begin{figure}
\includegraphics[height=3.3in,width=3.3in]{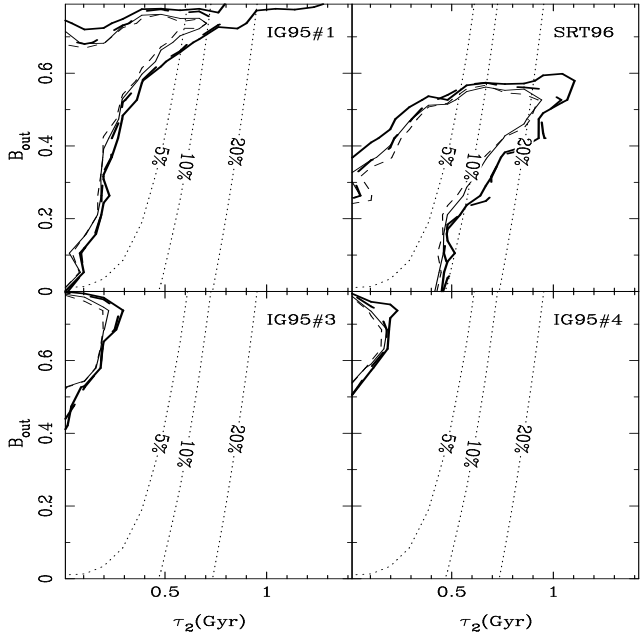}
\caption{Probability maps at the $90\%$ and $95\%$ confidence levels in the
($\tau_2$,$\bout$) parameter space. The same notation as in 
Figure~\ref{fig:t1t2} is used.}
\label{fig:t2b}
\end{figure}


\begin{table}
\caption{Best fit parameters (All [Fe/H]; 90\% confidence levels)}
\label{tab:all}
\begin{center}
\begin{tabular}{c|cccc}
FIELD & Model & $\tau_1$/Gyr & $\tau_2$/Gyr & $\bout$\\
\hline\hline
IG95/\#1 & C10z5 & $<0.18$ & $<0.28$ & $0.43^{+0.10}_{-0.13}$\\
         &  C5z5 & $<0.31$ & $<0.44$ & $0.43^{+0.11}_{-0.11}$\\
         & C10z2 & $<0.17$ & $<0.27$ & $0.43^{+0.09}_{-0.15}$\\

SRT96    & C10z5 & $<0.35$ & $<0.45$ & $0.11^{+0.10}_{-0.11}$\\
         &  C5z5 & $<1.14$ & $<0.74$ & $0.11^{+0.12}_{-0.11}$\\
         & C10z2 & $<0.28$ & $<0.46$ & $0.11^{+0.08}_{-0.11}$\\

IG95/\#3 & C10z5 & $<0.09$ & $<0.10$ & $0.64^{+0.04}_{-0.11}$\\
         &  C5z5 & $<0.13$ & $<0.18$ & $0.64^{+0.05}_{-0.10}$\\
         & C10z2 & $<0.09$ & $<0.10$ & $0.64^{+0.04}_{-0.11}$\\

IG95/\#4 & C10z5 & $<0.09$ & $<0.09$ & $0.59^{+0.03}_{-0.12}$\\
         &  C5z5 & $<0.10$ & $<0.13$ & $0.59^{+0.04}_{-0.12}$\\
         & C10z2 & $<0.09$ & $<0.09$ & $0.59^{+0.03}_{-0.14}$\\

Z03      & C10z5 & $<0.36$ & $<0.66$ & $0.59^{+0.05}_{-0.10}$\\
         &  C5z5 & $<1.15$ & $<1.02$ & $0.59^{+0.05}_{-0.12}$\\
         & C10z2 & $<0.36$ & $<0.61$ & $0.53^{+0.05}_{-0.05}$\\
\end{tabular}
\end{center}
\end{table}

\begin{figure}
\includegraphics[height=3.3in,width=3.3in]{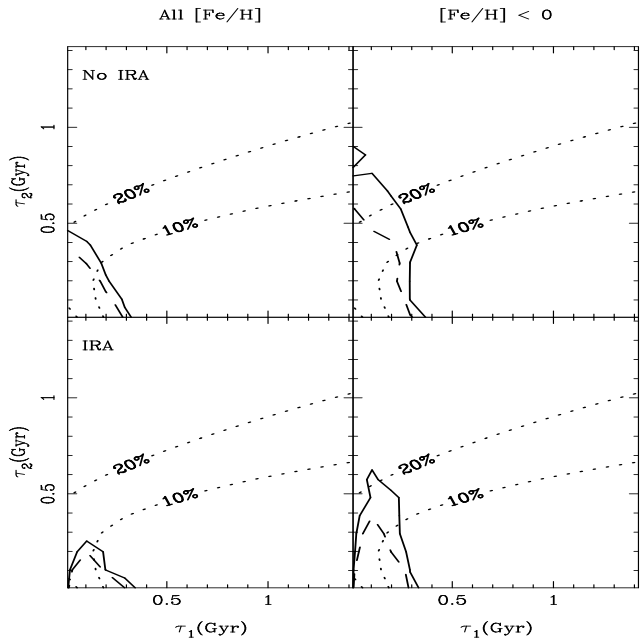}
\caption{The instantaneous recycling approximation (IRA) 
assumes stars have either zero or infinite lifetimes
depending on whether their masses are larger or smaller
than some characteristic mass scale. The bottom panels
show the probability maps when comparing the observed
metallicity distributions with a model which imposes IRA.
We have also set to zero the contribution from type~Ia
SNe. The top panels show the results from the full model
presented in this paper. The figure also shows the
difference when including stars with supersolar metallicity
to the analysis ({\sl left}). The solid (dashed) contours
represent the 90 (95) \% confidence level.
}
\label{fig:ira}
\end{figure}

Given that the infall timescales predicted by the models are rather
short, we decided to explore the validity of this result
by performing the same test on a model which adopts the
instantaneous recycling approximation (IRA; Tinsley 1980). In the IRA,
the stellar lifetimes are assumed to be either zero or infinity,
depending on whether the stellar mass is above or below some mass
threshold, respectively.  In that case, only stars with masses above
the threshold will contribute to the enrichment of the subsequent
stellar generations.  Furthermore, in this approximation we assumed
type~Ia supernovae do not contribute to the chemical
enrichment. Figure~\ref{fig:ira} shows the result. The probability
maps of our full model are shown in the top panels, when comparing
stars with subsolar metallicities ({\sl right}), or the full sample
({\sl left}) of field IG95/\#1. The bottom panels show the result in
the IRA. The infall timescales obtained are similar, with $\tau_1$ or
$\tau_2\simgt 0.5$~Gyr ruled out at more than the 90~\% confidence
level. Hence, the main conclusion of this paper -- namely that star
formation timescales $\simgt 1$~Gyr are ruled out by the distribution
of the metallicities of K giant bulge stars -- is a rather robust
statement which does not depend critically on the details of the 
adopted stellar lifetimes.

\subsection{Linear vs Non-linear Schmidt laws}

\begin{figure}
\includegraphics[width=3.4in]{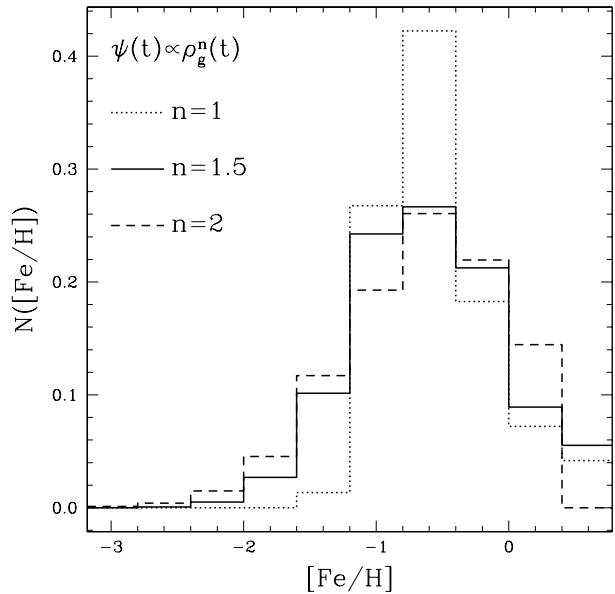}
\caption{Metallicity distribution predicted for 
three different exponents of the Schmidt law
relating the star formation rate with the gas density.
The model corresponds to $\tau_1=\tau_2=0.1$~Gyr;
$\bout =0.5$; $\ceff =10$. Notice that a linear
law gives a much sharper histogram  
since star formation takes longer for the same amount
of gas, compared to a non-linear law with $n>1$.
}
\label{fig:schmidt}
\end{figure}

One could expect that the short timescales obtained in the model
presented here could vary significantly when changing the dependence
between gas density and star formation. We have used throughout this
paper a Schmidt law with an exponent of $1.5$
(equation~\ref{eq:schmidt}).  Figure~\ref{fig:schmidt} shows the
predicted metallicity distribution when the exponent is changed 
first to a linear, and secondly to a quadratic, Schmidt law. The
difference between the predictions of our fiducial $n=1.5$
model and those with  $n=2$ is not large. However, a linear
law ($n=1$) gives a significantly sharper histogram, to be expected
since a linear law extends the period of star formation for the same
amount of gas and star-formation efficiency.  Longer star formation 
timescales (for fixed infall timescales) 
generate more stars with higher metallicity, thereby making the
histogram narrower -- much in the same way as more extended infall 
for a fixed star formation timescale.  Therefore, our conclusion 
regarding the need for short star formation timescales is independent 
of the exponent used in the Schmidt law.  Furthermore, a linear dependence 
would require even shorter infall timescales~!

\subsection{Bulge Spectrophotometry}

\begin{figure}
\includegraphics[width=3.4in]{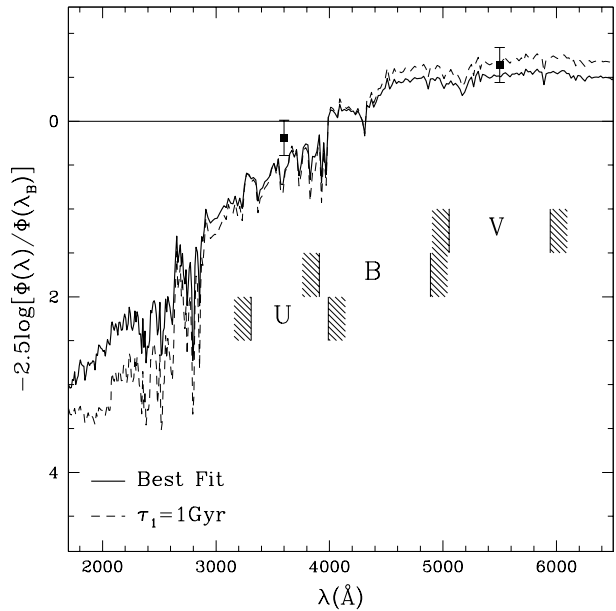}
\caption{Spectral energy distribution (sed) obtained by convolving
the star formation history for the best fit to Ibata \& Gilmore
field IG95\#1 with the stellar population synthesis models of Bruzual 
\& Charlot (in preparation). The model corresponds 
to $\tau_1=\tau_2=0.05$~Gyr; $\bout=0.5$. The vertical lines give 
the central positions of the $U$, $B$, and $V$ passbands. 
A second sed with a longer timescale ($\tau_1=1$~Gyr) is also shown 
(dashed line). The larger population of stars
with a higher metallicity in the $\tau_1=1$~Gyr model is responsible 
for the redder colours. The dots correspond to the average and 
standard deviation of the Sbc bulge sample of Gadotti \& dos~Anjos (2001).
A comparison of the solid and dashed lines shows that it is very difficult
to determine SFH parameters from integrated spectroscopy alone. The 
distribution of stellar metallicities (as presented in this
paper) is a much more sensitive discriminator.
}
\label{fig:sed}
\end{figure}

We can use the star formation history that gives the best fit to
the observed histograms and convolve it with simple stellar populations
over the range of ages and metallicities predicted by the model. We 
used the latest version of the population synthesis models of 
Bruzual \& Charlot (1993; priv.~comm.) in order to generate an 
integrated spectral energy 
distribution (sed) as shown in figure~\ref{fig:sed}, which 
corresponds to $\tau_1=\tau_2=0.2$~Gyr; $\bout=0.5$. The sed has been 
normalized to the flux in the $B$-band. The resulting sed gives 
colours $U-V=1.13$; $V-K=2.86$. For comparison purposes we have
considered another model with a longer infall timescale 
($\tau_1=1$~Gyr; dashed line). The colours of this model ---
$U-V=1.34$; $V-K=2.99$ --- are redder because of the higher 
metallicities caused by longer infall timescales. The dots
with error bars are the average and standard deviation of the
$U-B$ and $B-V$ colours from the sample of 257 bulges from 
Sbc spirals of Gadotti \& dos~Anjos (2001). Our predicted 
$B-V= 0.4$ colour is compatible with the average 
and standard deviation of the observed colour $B-V=0.64\pm 0.20$ 
(using a subsample of bulges with negative colour gradients).
However, our predicition for $U-B=0.6$ is significantly
bluer than the observed $U-B=0.19\pm 0.20$. This may be 
due to uncertainties in either the chemical enrichment model
or the population synthesis models considered. 
Figures~\ref{fig:t1b} and \ref{fig:t2b} show that the value 
of $\bout$ is rather uncertain and this has a strong 
effect in the final colours. Furthermore, 
the discrepancy towards bluer observed colours may be caused 
by a systematic disk contamination in the observations. Even though
less than 5\% of the $257$ bulges observed by Gadotti \& dos~Anjos (2001)
have colours $U-B<0.5$, a comparison with the colours of early-type
galaxies $U-B\sim 0.6-0.7$ (e.g. Gonz\'alez 1993) shows that our
model predictions are compatible with the photometry of spheroids.

Figure~\ref{fig:sed} illustrates the challenging task of estimating
infall timescales from $UBV$ broadband photometry alone. Only the spectral
window around $2000$\AA\  could be useful in order to rule out formation
timescales longer than $1$~Gyr. In that spectral region, the differences
can be as large as $1$~magnitude although the analysis would be hindered
by the many uncertainties behind the model of star formation and chemical 
enrichment as well as by the uncertainties in the model predictions from
population synthesis models. The direct comparison of stellar metallicites
is thereby a much more powerful technique to infer the star 
formation history of a bulge, stellar cluster, or galaxy.


\subsection{Simple Model and the duration of star formation}

The broad tail of the metallicity distribution at low values
of [Fe/H] could imply that the Galactic bulge can be reasonably fit by the
so-called Simple Model (e.g. Pagel 1997), which assumes a closed
box system and the instantaneous recycling approximation. The beauty
of the model lies in the ability to generate a metallicity distribution
which is completely independent of the star formation history.
The histogram is thereby degenerate with respect to the duration of the 
star formation process. It only depends on the total stellar yield $y$ 
as a scale factor in the following way:
\be
\frac{dM_\star}{d\log Z} \propto (Z/y)\exp (-Z/y),
\ee
where $Z$ is the total metallicity and $M_\star$ is the stellar mass 
content. This distribution has a rather broad range of low-metallicity
stars, which has been the reason why a Simple Model was discarded
to explain the formation of the local Galactic disk as it generated
too many G-dwarf stars at low metallicities (e.g. Tinsley 1980).
However, the bulge distributions shown in figure~\ref{fig:data}
are broader and so, more compatible with this model.

\begin{figure}
\includegraphics[width=3.4in]{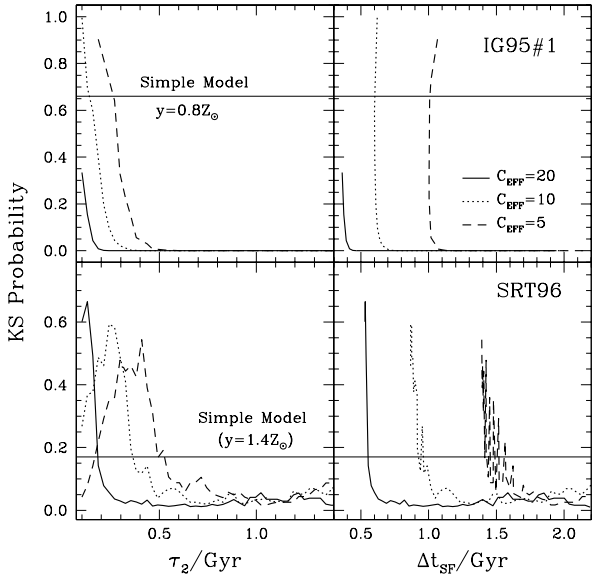}
\caption{Comparison of the Kolmogorov-Smirnov probability as a function
of $\tau_2$ ({\sl left}) or $\Delta t_{\rm SF}$ ({\sl right}). The latter
gives the time for the bulge to generate 75\% of the total stellar mass content.
Fields IG95\#1 ({\sl top}) and SRT96 ({\sl bottom}) 
are tested with a fixed value of $\bout = 0.5$ (IG95\#1) and $\bout=0$ (SRT96)
and $\tau_1=0.1$~Gyr, $z_F=5$ for both fields. Three star formation efficiencies
are chosen, namely $\ceff = 5$ (dashed), $10$ (dotted), and $20$ (solid).
The horizontal line gives the best fit for a Simple Model.
}
\label{fig:simple}
\end{figure}

We compared the best fits of a Simple Model with those from
our infall model in figure~\ref{fig:simple}. We plot the 
Kolmogorov-Smirnov probability when comparing fields IG95\#1 
({\sl top}) and SRT96 ({\sl bottom}) with our model for a fixed 
formation redshift: $z_F=5$, $\tau_1=0.1$~Gyr. The outflow parameter
was chosen to maximise the probability for each field. We take
$\bout=0.5$ for IG95\#1 and $\bout=0$ for SRT96.
Three different star formation efficiencies are considered,
namely  $\ceff=\{5, 10, 20\}$. The KS probability is shown 
as a function of $\tau_2$ ({\sl left}) or $\Delta t_{\rm SF}$ ({\sl right}),
where the latter is defined as the time lapse during which 75\% of the
total stellar mass content in the bulge is generated.
Only sub-solar metallicities are considered in the test.
The horizontal line gives the highest probability for a Simple Model,
varying the yield ($y$) in order to maximize the probability.
Hence, one can see that
our models give better fits than a Simple Model, so that a detailed
analysis of the metallicity distribution in the bulge can be used to 
set limits on the duration of star formation. All models shown in
the figure give star formation timescales shorter than $\sim 1.5$~Gyr.
Furthermore, the [Mg/Fe] enhancement observed in 
bulge stars (Rich \& McWilliam 2000) speaks in favour of models with 
a high $\ceff$ so that star formation timescales $\Delta t_{\rm SF}\simlt 0.5$~Gyr 
are expected.

\section{Conclusions}

We have explored a simple model describing the formation and
evolution of the stellar populations of the Galactic bulge using
a set of a few parameters. A comparison of the model predictions
with the observed [Fe/H] distribution of K giants in various
fields towards the bulge requires relatively short infall timescales
($\simlt 0.5$~Gyr) regardless of the field considered 
(Figure~\ref{fig:t1t2}). Our model compares the resulting star formation
history and subsequent enrichment with the distribution of stellar
metallicities. Within the model assumptions and other uncertainties,
one can relate the infall timescales with the duration of star formation.
Hence, formation timescales 
longer than $\simgt 1$~Gyr are ruled out at more than the 90~\%
confidence level regardless of the field, statistical test, 
or on whether stars with supersolar 
metallicities are excluded from the analysis (Figure~\ref{fig:ira}). 

Outflows during the formation of the bulge can also be estimated,
although this result is very strongly dependent on the (uncertain)
absolute stellar yields of Fe from type~II supernovae
and requires a precise calibration of the absolute metallicities
of the observed stars, including the effect of non-solar abundance
ratios.  Taking the derived distributions as face value, even if we
decided to track the total metallicity, $Z$, instead of the Fe content, 
the uncertainties in the stellar yields from intermediate mass stars
are still large enough so that the model estimates of the fraction of
gas and metals ejected in outflows are more qualitative than quantitative.
Nevertheless, we find outflows may be very significant in most
of the fields studied. Baade's window (SRT96) is the field in which high
outflow fractions are ruled out. On the other hand, the fields from IG95
favour a large amount of gas ejected in outflows.
In these fields, $\bout\sim 0.5-0.7$ give the best fit, and $\bout =0$ 
seems to be a very unlikely scenario even if the real stellar yields 
or the IMF are far from those adopted in this paper. This difference 
in outflow fraction is in the sense expected if outflow is inhibited in 
the deepest part of the potential well. Large outflow fractions are 
expected and predicted in some models that include estimates of the 
dynamical effects of feedback from  stars (e.g. Arimoto \& Yoshii 1987). 
It is worth noticing that these
values of $\bout$ imply a very significant amount of metals contributing
to the enrichment of the IGM (Renzini 1997), provided the gas escapes the 
overall Galaxy, rather than for example enriching the disk.

\begin{figure}
\includegraphics[width=3.4in]{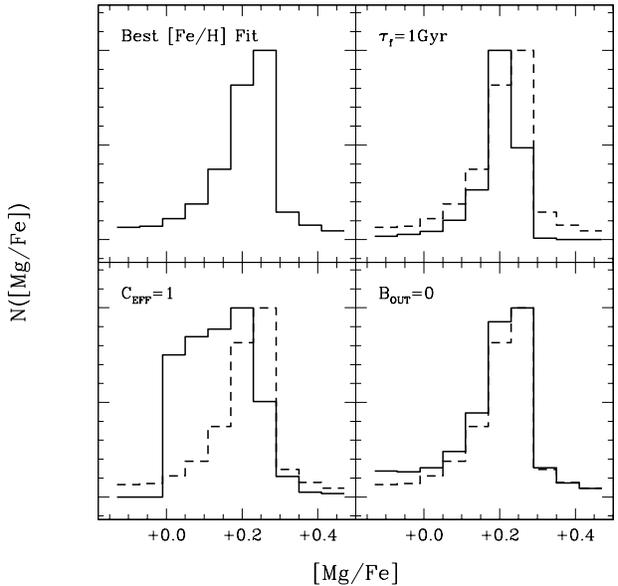}
\caption{Distribution of [Mg/Fe] abundances predicted by the models
shown in figure~\ref{fig:hist}. The histogram corresponding to the
best fit is shown in the top-left panel as well as in the three
remaining panels as a dashed line. Notice that a low star formation 
efficiency -- which generates a similar [Fe/H] as the best fit as
seen in figure~\ref{fig:hist} -- generates a very wide [Mg/Fe] distribution.
}
\label{fig:mgfe}
\end{figure}

The short timescales predicted by the analysis of the metallicity
distribution of bulge stars has a direct consequence on 
abundance ratios such as [Mg/Fe]. This ratio is a reasonably robust
indicator of the duration of star formation. Solar values are
achieved when extended star formation takes place, so that the debris
from SNeIa can be incorporated into subsequent generations of stars.
On the other hand, short-lived bursts such as the one we predict in
this paper, translates into enhanced [Mg/Fe] in most bulge stars. 
Figure~\ref{fig:mgfe}
shows this point. Analogously to figure~\ref{fig:hist}, we give the
model prediction for various choices of the parameters including
the set that gives the best fit. In this case the histogram of
[Mg/Fe] is shown. In most cases, the distribution is very similar,
peaked at [Mg/Fe]$\sim +0.25$ with an extended tail which
dies off at solar abundance ratios. Only the model with low star
formation efficiency ({\sl bottom left}) gives a significantly different
histogram. A low $\ceff$ implies a more extended period of star
formation, generating a broader histogram as more stars become polluted
by SNeIa ejecta. The abundance ratios observed by Rich \& McWilliam (2000) 
on a sample of bulge giants using Keck/HIRES rule out
models with low star formation efficiencies. 

The main outcome of this paper is that infall timescales
$\tau\simgt 1$~Gyr are ruled out by the {\sl observed} metallicity 
distribution of the Galactic bulge. Short star formation times are also
to be expected given the observed broad distribution of the metallicities
of bulge K giants. The preferred value stays around infall timescales 
$\tau_f\sim 0.1-0.2$~Gyr, which would correspond to star formation
timescales $\tau_{\rm SF}\simlt 1$~Gyr. 
 However, the differences between the iron-abundance distributions
from the different fields explored in this paper illustrate the
need for a uniform large-scale (IR) spectroscopic survey, with
radial velocity AND STAR COUNTS used as an aid to the statistical 
separation of bulge and foreground disk AND STELLAR HALO. 

A robust calibration of metallicities -- which should account for
variations in the [$\alpha$/Fe] abundance ratios -- is crucial in
the quantification of the amount of gas ejected in outflows
during the formation of the bulge. Furthermore, an accurate
distribution of abundance ratios such as [Mg/Fe] (figure~\ref{fig:mgfe}) 
would pose
strong constraints on the duration of the star formation burst
which gave way to the Galactic bulge.

Our technique is quite robust and complementary to
a comparison of the stellar colour-magnitude diagrams of old
populations. The latter can only constrain the star formation timescale
to a few Gyr due to the crowding of the isochrones at old ages. The model
presented here sets strong constraints for 
bulge formation in semi-analytical models, the latest of which predict 
formation timescales for bulges significantly longer than these
timescales (Abadi et al.~2002). Furthermore, secular evolution
models predict formation times which extend over many
dynamical timescales. Hence, formation scenarios over times 
$\sim 0.1-0.2$~Gyr, as presented here, would be in conflict with 
bulge formation from bar instabilities. Our results for the
buildup of the Galactic bulge
favour formation scenarios in which a strong starburst quickly
converts all gas available into stars in a few dynamical
timescales (Elmegreen 1999).

\section*{Acknowledgments}
We would like to thank the anonymous referee for his/her
comments and suggestions which have helped in 
clarifying the science presented in this paper.

\bsp

\label{lastpage}


\begin{thebibliography}{99}
\bibitem{ab02} Abadi, M.~G., Navarro, J.~F., Steinmetz, M.
  \& Eke, V.~R., 2002, astro-ph/0212282
\bibitem{ay87} Arimoto, N. \& Yoshii, Y., 1987, A\&A, 173, 23
\bibitem{bc93} Bruzual, A.~G. \& Charlot, S., 1993, ApJ, 405, 538
\bibitem{ead01} Ellis, R.~S., Abraham, R.~G. \& Dickinson, M.,
  2001, ApJ, 551, 111
\bibitem{el99} Elmegreen, B.~G., 1999, ApJ, 517, 103
\bibitem{fab85} Faber, S.~M., Friel, E.~D., Burstein, D. \&
  Gaskell, C.~M., 1985, ApJS, 57, 711
\bibitem{fg00} Feltzing, S. \& Gilmore, G., 2000, A\&A, 355, 945; 
  erratum 2001, A\&A, 369, 510
\bibitem{fs00a} Ferreras, I. \& Silk, J., 2000a, ApJ, 532, 193
\bibitem{fs00b} Ferreras, I. \& Silk, J., 2000b, MNRAS, 316, 786
\bibitem{fs01} Ferreras, I. \& Silk, J., 2001, ApJ, 557, 165
\bibitem{fhp} Fukugita, M., Hogan, C. \& Peebles, P.~J., 1998, ApJ, 503, 518
\bibitem{gd01} Gadotti, D.~A. \& dos Anjos, S., 2001, AJ, 122, 1298
\bibitem{jjg93} Gonz\'alez, J.~J., 1993, Ph.D. thesis, UC Santa Cruz
\bibitem{gr83} Greggio, L. \& Renzini, A., 1983, A\&A, 118, 217
\bibitem{har76} Hartwick, F.~D.~A., 1976, ApJ, 209, 418
\bibitem{ig95a} Ibata, R.~A. \& Gilmore, G.~F., 1995a, MNRAS, 275, 591 (IG95)
\bibitem{ig95b} Ibata, R.~A. \& Gilmore, G.~F., 1995b, MNRAS, 275, 605
\bibitem{ib84} Iben, I. \& Tutukov, A., 1984, ApJS, 54, 335
\bibitem{iwa99} Iwamoto, K., Brachwitz, F., Nomoto, K., Kishimoto, N., 
  Umeda, H., Hix, W.~R., Thielemann, F.-K., 1999, ApJS, 125, 439
\bibitem{kr02} Kuijken, K., \& Rich, R.~M., 2002, AJ, 124, 2054
\bibitem{ka96} Kauffmann, G., 1996, MNRAS, 281, 487
\bibitem{ken98} Kennicutt, R.~C., 1998, ApJ, 498, 541
\bibitem{lar74} Larson, R.~B., 1974, MNRAS, 169, 229
\bibitem{dlb75} Lynden-Bell, D., 1975, Vistas in Astronomy 19, 299
\bibitem{mc00} McNamara, D.~H., Madsen, J.~B., Barnes, J. \& Ericksen, B.~F.,
  2000, PASP, 112, 202
\bibitem{mc94} McWilliam, A. \& Rich, R.M. 1994, ApJS, 91, 749
\bibitem{mar03} Maraston, C., et al.~2003, A\&A in press (astro-ph/0209220)
\bibitem{mg86} Matteucci, F. \& Greggio, L., 1986, A\&A, 154, 279
\bibitem{mr01} Matteucci, F. \& Recchi, S., 2001, ApJ, 558, 351 
\bibitem{mrm99} Matteucci, F., Romano, D. \& Molaro, P., 1999, A\&A, 341, 458
\bibitem{mfg00} Moll\'a, M., Ferrini, F. \& Gozzi, G., 2000, MNRAS, 316, 345
\bibitem{mo84} Mould, J.~R., 1984, PASP, 96, 773
\bibitem{no97} Nomoto, K., Iwamoto, K. \& Kishimoto, N.,
  1997, Science, 276, 1378
\bibitem{nsh96} Norman, C.~A., Sellwood, J.~A. \& Hasan, H., 
  1996, ApJ, 462, 114
\bibitem{or95} Ortolani, S., Renzini, A., Gilmozzi, R., Marconi, G., 
  Barbuy, B., Bica, E. \& Rich, R.~M., 1995, Nature, 377, 701
\bibitem{pag97} Pagel, B.~E.~J. ``Nucleosynthesis and chemical
  evolution of galaxies'', 1997, Cambridge, p. 218
\bibitem{pel99} Peletier, R.~F., Balcells, M., Davies, R.~L., Andredakis, Y.,
  Vazdekis, A., Burkert, A. \& Prada, F., 1999, MNRAS, 310, 703
\bibitem{psr00} Proctor, R.~N., Sansom, A.~E. \& Reid, I.~N., 
  2000, MNRAS, 311, 37
\bibitem{ra91} Raha, N., Sellwood, J.~A., James, R.~A. \& Kahn, F.~D.,
  1991, Nature, 352, 411
\bibitem{re97} Renzini, A. 1997, ApJ, 488, 35
\bibitem{ri90} Rich, R.~M., 1990, ApJ, 362, 604
\bibitem{rmcw00} Rich, R.~M. \& McWilliam, A., 2000, Proc. SPIE, 4005, 150
\bibitem{srt96} Sadler, E.~M., Rich, R.~M. \& Terndrup, D.~M., 
  1996, AJ, 112, 171 (SRT96)
\bibitem{scs93} Salaris, M., Chieffi, A. \& Straniero, O., 
  1993, ApJ, 414, 580
\bibitem{sal55} Salpeter, E.~E. 1955, ApJ, 121, 161
\bibitem{sc86} Scalo, J.~M. 1986, Fund.Cosm.Phys. 11, 1
\bibitem{scha92} Schaller, G., Schaerer, D., Maeder, A. \& Meynet, G.
  1992, A\&AS, 96, 269
\bibitem{sch63} Schmidt, M. 1959, ApJ 129, 243
\bibitem{shap83} Shapiro, S.~L. \& Teukolsky, S.~A. 1983, {\sl Black Holes,
  White Dwarfs and Neutron Stars}, New York, Wiley-Interscience
\bibitem{th96} Thielemann, K.~F., Nomoto, K. \& Hashimoto, M.
  1996, ApJ, 460, 408
\bibitem{ti75} Tinsley, B.~M. 1975, ApJ, 197, 159
\bibitem{ti80} Tinsley, B.~M. 1980, Fund.Cosm.Phys., 5, 287
\bibitem{ty80} Tutukov, A.~V. \& Yungelson, L.~R. 1980, in {\sl Close
  binary stars}, IAU symp. no. 88, eds. M.~J. Plavec, 
  D.~M. Popper and R.~K. Ulrich, Reidel, Dordrecht, p.15
\bibitem{vl03} van Loon, J.~Th., et al.~2003, MNRAS, 338, 857 
\bibitem{ww95} Woosley, S. \& Weaver, T. 1995, ApJS, 101, 181
\bibitem{zoc03} Zoccali, M., et al., 2003, A\&A, 399, 931
\end{thebibliography}
\end{document}